\documentclass[groupedaddress]{revtex4}

\usepackage{amsmath}

\begin{document}

\title{
Eulerian Derivation of the Coriolis Force
}

\author{Akira Kageyama}
\email{kage@jamstec.go.jp}
\author{Mamoru Hyodo}
\email{hyodo@jamstec.go.jp}
\affiliation{
The Earth Simulator Center,\\
Japan Agency for Marine-Earth Science and Technology,\\
Yokohama 236-0001, Japan.
}

\begin{abstract}
In textbooks of geophysical fluid dynamics,
the Coriolis force and the centrifugal force in a rotating fluid system
are derived by making use of 
the fluid parcel concept.
In contrast to this intuitive derivation to the apparent forces,
more rigorous derivation
would be useful not only for the pedagogical purpose, 
but also for the applications to 
other kinds of rotating geophysical systems rather than the fluid.
The purpose of this paper is to show a general procedure
to derive the transformed equations in the rotating frame of reference
based on the local Galilean transformation and rotational
coordinate transformation of field quantities.
The generality and usefulness of this Eulerian approach is demonstrated
in the derivation of apparent forces in rotating fluids
as well as the transformed 
electromagnetic field equation in the rotating system.
\end{abstract}

\maketitle

%----------------------------
\section{Introduction}
%----------------------------
%
%
In textbooks of geophysical 
fluid dynamics~\citep[e.g.,][]{
pedlosky_1979,
stommel_1989,
cushman-roisin_1994,
salmon_1998}
and educational web sites~\citep[e.g.,][]{price_2004},
the apparent forces---the Coriolis force 
and the centrifugal force---are derived 
with the help of the framework of classical mechanics of a point particle;
it is first shown that the time derivative of a vector $\mathbf{A}$
is written as
$
(d\mathbf{A}/dt)_I =
(d\mathbf{A}/dt)_R + \mathbf{\Omega}\times\mathbf{A},
$
where 
$\mathbf{\Omega}$ is a constant angular velocity of the rotating frame of reference;
$I$ and $R$ stand for the inertial and rotating frames, respectively.
The above relation for vector $\mathbf{A}$
 is applied to a fluid parcel's position $\mathbf{r}$ and then
to its velocity $\mathbf{u}$,
leading to the relation
\begin{equation} \label{eq:0954}
(d\mathbf{u}_I/dt)_I = (d\mathbf{u}_R/dt)_R + 2\mathbf{\Omega}\times\mathbf{u}_R
+ \mathbf{\Omega}\times(\mathbf{\Omega}\times\mathbf{r}).
\end{equation}
Assuming that $(d\mathbf{u}_I/dt)_I$ equals to the local force
acting per unit mass on a fluid parcel,
the apparent forces in the rotating frame are derived.

The above derivation
can be called as a Lagrangian approach since
it exploits the concept of the fluid parcel.
This Lagrangian derivation seems to be a standard style
not only in the field of geophysical fluid dynamics, but also in
more general fluid dynamics~\citep[e.g.,][p.140]{bachelor_1970}.

The conventional Lagrangian derivation
is ingenious and simple enough for introductory courses.
But, why do we have to use the (Lagrangian) fluid parcel concept
when we just want to derive the (Eulerian) field equation in the rotating frame?
It should be possible to derive the apparent forces
by a straightforward, rotating coordinate
transformation of field quantities and the equation.
The purpose of this paper is to introduce such an Eulerian derivation
of the apparent forces
that can be contrasted with the conventional Lagrangian derivation.

There are three advantages of 
the Eulerian approach shown in this paper
compared to the conventional Lagrangian approach.
Firstly, it is general.
The Eulerian transformation is derived
for any vector field [eqs.~(\ref{eq:c0013}) and~(\ref{eq:c0021})].
Therefore, in addition to the fluid system,
it can be used to derive, for example, 
the Maxwell's equations in a rotating frame of reference
in which the fluid parcel concept is invalid. 
(The Lagrangian approach does not work unless one could define
proper working Lagrangian vector like $\mathbf{A}$ for the electromagnetic field.)

Secondly, physical meaning of the Eulerian derivation is clear.
The Eulerian transformation of a vector field is composed of
the local Galilean transformation and rotational transformation,
as we will see in eq.~(\ref{eq:c0013}).
The transformation of the time derivative of a vector field
[shown in eq.~(\ref{eq:c0021})] is also
described by the local Galilean transformation and
rotational transformation.

Thirdly, it is mathematically rigorous.
The key of the simplicity of the conventional Lagrangian derivation is
eq.~(\ref{eq:0954}).
But note that the expression of
$d\mathbf{u}/dt$ is actually an abbreviated form,
when it is finally applied to the fluid equation,
of rather complicated terms;
$\partial\mathbf{u}/\partial t + (\mathbf{u}\cdot\nabla)\mathbf{u}$.
Note that the second term is nonlinear of $\mathbf{u}$.
If one substitutes
$(d\mathbf{u}_I/dt)_I=\partial\mathbf{u}_I/\partial t
 + (\mathbf{u}_I\cdot\nabla_I)\mathbf{u}_I$,
 and 
$(d\mathbf{u}_R/dt)_R=\partial\mathbf{u}_R/\partial t
 + (\mathbf{u}_R\cdot\nabla_R)\mathbf{u}_R$,
into the left-hand and right-hand sides of eq.~(\ref{eq:0954}), respectively,
the covered complexity of the equation becomes apparent
that requires a mathematical proof.

In our opinion, the conventional Lagrangian derivation
of the apparent forces is intuitive and simple, 
while the Eulerian derivation shown here is rigorous and straightforward.
These two derivations would be regarded as
a complemental approaches to the understanding of the 
apparent forces including the Coriolis force.

The authors could not find the Eulerian derivation 
of this kind
in textbooks on fluid and geophysical fluid dynamics.

%----------------------------
\section{Vector fields in a rotating frame of reference}
%----------------------------
%
In order to derive the general expression of a vector field in a rotating frame of reference,
we start from the Galilean transformation of a vector field 
between two inertial frames.
Let $L_I$ and $L'_I$ be inertial frames with
relative velocity $\mathbf{V}$;
\begin{equation} \label{eq:c0001}
  \mathbf{x}' = \mathbf{x} - \mathbf{V} t,
\end{equation}
where $\mathbf{x}$ and $\mathbf{x}'$ are coordinates in $L_I$ and $L'_I$, respectively.
When a vector field $\mathbf{a}(\mathbf{x},t)$ is defined in $L_I$,
it is observed in $L'_I$ as
\begin{equation} \label{eq:c0002}
  \mathbf{a}'(\mathbf{x}',t) = G^\mathbf{V}\, \mathbf{a}(\mathbf{x},t),
\end{equation}
where $G^\mathbf{V}$ is a Galilean transformation operator.
For example, the transformation of a fluid flow $\mathbf{u}(\mathbf{x},t)$ is given by
\begin{equation} \label{eq:c0003}
  \mathbf{u}'(\mathbf{x}',t) = G^\mathbf{V}\, \mathbf{u}(\mathbf{x},t) \equiv \mathbf{u}(\mathbf{x},t) - \mathbf{V}.
\end{equation}
Other examples of the Galilean transformation operator $G^\mathbf{V}$ 
are for the magnetic field $\mathbf{B}$ and 
the electric field $\mathbf{E}$;
\begin{equation} \label{eq:c0004}
  \mathbf{B}'(\mathbf{x}',t) = G^\mathbf{V}\, \mathbf{B}(\mathbf{x},t) \equiv \mathbf{B}(\mathbf{x},t),
\end{equation}
\begin{equation} \label{eq:c0005}
  \mathbf{E}'(\mathbf{x}',t) = G^\mathbf{V}\, \mathbf{E}(\mathbf{x},t) 
                                          \equiv \mathbf{E}(\mathbf{x},t) + \mathbf{V}\times \mathbf{B}(\mathbf{x},t).
\end{equation}
These transformations are derived 
from the
Lorentz transformation
in the limit of $V \ll c$, where $c$ is the speed of light.
When a vector field $\mathbf{F}$ is a function of a vector field $\mathbf{a}$, $\mathbf{F} = \mathbf{F}(\mathbf{a})$,
its transformation is given by
\begin{equation} \label{eq:c2044}
  \mathbf{F}'
     = G^\mathbf{V}\,\mathbf{F}(\mathbf{a},t)
     \equiv \mathbf{F}(G^{\mathbf{V}}\mathbf{a},t).
\end{equation}
For example,
when $\mathbf{F} = (\mathbf{u}\cdot\nabla)\mathbf{u}$,
\begin{equation} \label{eq:c2045}
  (\mathbf{u}'\cdot\nabla')\mathbf{u}'
 = G^{\mathbf{V}}\,   (\mathbf{u}\cdot\nabla)\mathbf{u}
 \equiv [(\mathbf{u}(\mathbf{x},t)-\mathbf{V})\cdot \nabla]
 (\mathbf{u}(\mathbf{x},t)-\mathbf{V}),
\end{equation}
where we have used the equivalence of the operators
$\nabla$ and $\nabla'$ defined for two inertial frames $L$ and $L'$.

Let $\hat{L}_R$ be a rotating frame of reference with constant angular velocity $\mathbf{\Omega}$
with respect to $L_I$.
For simplicity, we suppose that $\hat{L}_R$ and $L_I$ share 
the same origin and $z$ axis,
and $\hat{L}_R$ is rotating around the $z$ axis; $\mathbf{\Omega} = (0,0,\Omega)$.
The coordinates $\mathbf{x}$ and $\hat{\mathbf{x}}$ of a point 
observed in $L_I$ and $\hat{L}_R$ are related as
\begin{equation} \label{eq:c0006}
  \hat{\mathbf{x}} = R^{\Omega t} \mathbf{x},
\end{equation}
where
$R^{\Omega t}$ denotes the rotational transformation with matrix expression
\begin{equation}  \label{eq:c0007}
  R^{\Omega t} = 
   \left(
     \begin{array}{ccc}
         \cos{\Omega t} & \sin{\Omega t} & 0 \\
         -\sin{\Omega t} &  \cos{\Omega t} & 0 \\
           0   &   0  & 1 
     \end{array}
  \right).
\end{equation}

Suppose a point $P$ at 
coordinates $\hat{\mathbf{x}}$ that is fixed in the rotating frame $\hat{L}_R$.
The point $P$ is observed as a circular trajectory in $L_I$.
Let $\mathbf{x}_t$ and $\mathbf{x}_{t+\Delta t}$ be two positions of $P$  
in the inertial frame $L_I$
at two successive time steps $t$ and $t+\Delta t$.
Equation~(\ref{eq:c0006}) reads
\begin{equation}  \label{eq:c0008}
  \hat{\mathbf{x}} = R^{\Omega (t+\Delta t)} \mathbf{x}_{t+\Delta t} = R^{\Omega t} \mathbf{x}_{t}.
\end{equation}
Since the inverse transformation of the 
rotation $R^{\Omega \Delta t}$ is  $R^{-\Omega \Delta t}$,
we obtain
\begin{equation}  \label{eq:c0009}
   \mathbf{x}_{t+\Delta t} = R^{-\Omega \Delta t} \mathbf{x}_{t}.
\end{equation}
By expanding eq.~(\ref{eq:c0009}) with respect to $\Delta t$  and ignoring  $O[(\Delta t)^2]$, we get
\begin{equation}  \label{eq:c0010}
   \mathbf{x}_{t+\Delta t} = \mathbf{x}_t + \Delta t \mathbf{\Omega}\times \mathbf{x}_t.
\end{equation}
Eq.~(\ref{eq:c0010}) means that
the point $P$  at position $\mathbf{x}$ at time $t$ is moving with 
instantaneous velocity $\mathbf{\Omega}\times\mathbf{x}$.

Suppose a vector field $\mathbf{a}(\mathbf{x},t)$ defined in the inertial frame $L_I$.
If another inertial frame $L'_I$ is co-moving
with point $P$ with velocity $\mathbf{\Omega}\times\mathbf{x}_P$,
where $\mathbf{x}_P$ is the $P$'s coordinates in $L_I$,
then the vector observed in $L'_I$  is
described by (instantaneous) Galilean transformation [eq.~(\ref{eq:c0002})], as
\begin{equation}  \label{eq:c0011}
   \mathbf{a}'(\mathbf{x}'_P,t) = G^\mathbf{V} \mathbf{a}(\mathbf{x}_P,t), \ \ \ \hbox{with}\  \mathbf{V} = \mathbf{\Omega}\times\mathbf{x}_P.
\end{equation}
The vector components of $\mathbf{a}'(\mathbf{x}'_P,t)$ 
in $L'_I$ 
and $\hat{\mathbf{a}}(\hat{\mathbf{x}}_P,t)$
in the rotating frame $\hat{L}_R$ are related 
by the same rotation matrix $R^{\Omega t}$ appeared in eq.~(\ref{eq:c0006});
\begin{equation}  \label{eq:c0012}
   \hat{\mathbf{a}}(\hat{\mathbf{x}}_P,t) = R^{\Omega t} G^\mathbf{V} \mathbf{a}(\mathbf{x}_P,t).
\end{equation}
Applying the above local Galilean transformation for the specific point $P$
to every position $\hat{\mathbf{x}}$ in $\hat{L}_R$, we obtain
\begin{equation}  \label{eq:c0013}
   \hat{\mathbf{a}}(\hat{\mathbf{x}},t) 
     = R^{\Omega t}\, G^\mathbf{\mathbf{\Omega}\times\mathbf{x}}\,  \mathbf{a}(\mathbf{x},t).
\end{equation}

For example, from eq.~(\ref{eq:c0003}),
the fluid flow $\mathbf{u}(\mathbf{x},t)$ in $L_I$
is transformed into $\hat{\mathbf{u}}(\hat{\mathbf{x}},t)$
in $\hat{L}_R$ as
\begin{equation}  \label{eq:c0014}
   \hat{\mathbf{u}}(\hat{\mathbf{x}},t) 
     = R^{\Omega t}\, 
        \left\{ 
          \mathbf{u}(\mathbf{x},t) - \mathbf{\Omega}\times\mathbf{x}
        \right\}
     =  R^{\Omega t}\, 
        \left\{
          \mathbf{u}(\mathbf{x},t) + \mathbf{x}\times\mathbf{\Omega}
        \right\}.
\end{equation}
Similarly, from eqs.~(\ref{eq:c0004}) and (\ref{eq:c0005}), we obtain the
transformation formulae for the magnetic field and the electric field:
\begin{equation}  \label{eq:c0015}
   \hat{\mathbf{B}}(\hat{\mathbf{x}},t) 
     =  R^{\Omega t}\, 
          \mathbf{B}(\mathbf{x},t),
\end{equation}
and
\begin{equation}  \label{eq:c0016}
   \hat{\mathbf{E}}(\hat{\mathbf{x}},t) 
     =  R^{\Omega t}\, 
        \left\{
          \mathbf{E}(\mathbf{x},t) + (\mathbf{\Omega}\times\mathbf{x})\times\mathbf{B}
        \right\}.
\end{equation}

To derive the fluid equation in the rotating frame $\hat{L}_R$,
we need to calculate the transformation of the flow's nonlinear term
$(\mathbf{u}\cdot\nabla)\mathbf{u}$;
\begin{align} \label{eq:c2058}
&  (\hat{\mathbf{u}}\cdot\hat{\nabla})\hat{\mathbf{u}}  \nonumber\\
& = R^{\Omega t}\, G^{\mathbf{\Omega}\times\mathbf{x}} (\mathbf{u}\cdot\nabla)\mathbf{u} \nonumber\\
& =  R^{\Omega t} 
   [
     \left\{(\mathbf{u}+\mathbf{x}\times\mathbf{\Omega})\cdot
     \nabla\right\}
      (\mathbf{u}+\mathbf{x}\times\mathbf{\Omega})
   ] \ \ \ \hbox{[cf. eq.~(\ref{eq:c2045})]} \nonumber\\
& = 
   R^{\Omega t} 
   [
   \{
        (\mathbf{u}+\mathbf{x}\times\mathbf{\Omega})\cdot
     \nabla 
   \} \mathbf{u}
     +
     \{
     (\mathbf{u}+\mathbf{x}\times\mathbf{\Omega})\cdot
      \nabla
     \} (\mathbf{x}\times\mathbf{\Omega})
    ] \nonumber\\
& = 
   R^{\Omega t} 
   [
   \{
        (\mathbf{u}+\mathbf{x}\times\mathbf{\Omega})\cdot
     \nabla
   \} \mathbf{u}
     +
     (\mathbf{u}+\mathbf{x}\times\mathbf{\Omega})
     \times\mathbf{\Omega}
    ] \ \ \ \hbox{[cf. (\ref{eq:c2249})]}\nonumber\\
& = 
   R^{\Omega t} 
   [
       (\mathbf{u}\cdot\nabla)\mathbf{u}
       +\{(\mathbf{x}\times\mathbf{\Omega})\cdot\nabla
         \}\mathbf{u}
     +
     \mathbf{u}\times\mathbf{\Omega}
     +(\mathbf{x}\times\mathbf{\Omega})\times\mathbf{\Omega}
    ].
\end{align}
Here, we have used the following equation for any vector $\mathbf{a}$
\begin{equation} \label{eq:c2249}
   (\mathbf{a}\cdot\nabla)(\mathbf{x}\times\mathbf{\Omega})
    = \mathbf{a}\times\mathbf{\Omega},
\end{equation}
with $\mathbf{a}=\mathbf{u}+\mathbf{x}\times\mathbf{\Omega}$.
In general, 
the transformation of a vector field $\mathbf{F}$ that is 
a function of a vector field $\mathbf{a}$, 
$\mathbf{F}=\mathbf{F}(\mathbf{a},t)$, is given 
from eq.~(\ref{eq:c2044}) as
\begin{equation}  \label{eq:c2307}
   \hat{\mathbf{F}}(\hat{\mathbf{a}},t) 
     = R^{\Omega t}\, G^\mathbf{\mathbf{\Omega}\times\mathbf{x}}\,  \mathbf{F}(\mathbf{a},t)
     = R^{\Omega t}\, \mathbf{F}(G^\mathbf{\mathbf{\Omega}\times\mathbf{x}}\,  \mathbf{a},t).
\end{equation}

The next step is to derive the transformation of the time derivative of a vector field, 
$\partial \hat{\mathbf{a}}(\hat{\mathbf{x}},t)/\partial t$,
where the partial derivative should be taken with fixed coordinates $\hat{\mathbf{x}}$
in the rotating frame of reference $\hat{L}_R$:
\begin{equation} \label{eq:c0017}
\frac{\partial \hat{\mathbf{a}}}{\partial t}(\hat{\mathbf{x}},t) 
= 
   \lim_{\Delta t\rightarrow 0}
     \frac{
      \hat{\mathbf{a}}(\hat{\mathbf{x}},t+\Delta t) - \hat{\mathbf{a}}(\hat{\mathbf{x}},t) 
      }
      {\Delta t},
\end{equation}
where, from eq.~(\ref{eq:c0013}),
\begin{equation} \label{eq:c0018}
  \hat{\mathbf{a}}(\hat{\mathbf{x}},t+\Delta t) 
   =   R^{\Omega (t+\Delta t)}\,
       G^{\mathbf{\Omega}\times\mathbf{x}_{t+\Delta t}}\, 
        \mathbf{a}(\mathbf{x}_{t+\Delta t},t+\Delta t),
\end{equation}
and
\begin{equation} \label{eq:c0019}
  \hat{\mathbf{a}}(\hat{\mathbf{x}},t) 
    = R^{\Omega t}\,
       G^{\mathbf{\Omega}\times\mathbf{x}_t}\, 
        \mathbf{a}(\mathbf{x}_t,t).
\end{equation}
Substituting eq.~(\ref{eq:c0010}) into 
$\mathbf{a}(\mathbf{x}_{t + \Delta t},t+\Delta t)$ 
and expanding it with respect to $\Delta t$ to the first order,
\begin{align} \label{eq:c0020}
&  \mathbf{a}(\mathbf{x}_{t+\Delta t},t+\Delta t) \nonumber\\
&   = 
    \mathbf{a}(\mathbf{x}_t,t)
    + \Delta t\, \left[(\mathbf{\Omega}\times \mathbf{x}_t)\cdot \nabla\right]\, \mathbf{a}(\mathbf{x}_t,t)
    + \Delta t\, \frac{\partial \mathbf{a}}{\partial t}(\mathbf{x}_t,t).
\end{align}
From eqs.~(\ref{eq:c0017})--(\ref{eq:c0020}) with the aid of
the following relations
\begin{equation} \label{eq:c0022}
  \lim_{\Delta t\rightarrow 0}
  \frac{R^{\Omega \Delta t}\, \mathbf{a} - \mathbf{a}}{\Delta t}
 = 
    \mathbf{a} \times \mathbf{\Omega},
\end{equation}
and
\begin{equation} \label{eq:d0022}
  G^{\mathbf{\Omega}\times\mathbf{x}_{t+\Delta t}}
 = 
    G^{\mathbf{\Omega}\times\mathbf{x}_t + \Delta t \mathbf{\Omega}\times(\mathbf{\Omega}\times\mathbf{x}_t)}
 =    
    G^{\mathbf{\Omega}\times\mathbf{x}_t}(1+O[\Delta t]),
\end{equation}
we get
\begin{equation}    \label{eq:c0021}
\frac{\partial \hat{\mathbf{a}}}{\partial t}(\hat{\mathbf{x}},t) 
= 
   R^{\Omega t}\, 
   [
       \frac{\partial}{\partial t}
        + (\mathbf{\Omega}\times\mathbf{x})\cdot \nabla
        - \mathbf{\Omega} \times
   ]
   \,        
   G^{\mathbf{\Omega}\times\mathbf{x}}\,
   \mathbf{a}(\mathbf{x},t).
\end{equation}
Here we have used $\mathbf{x}$ instead of $\mathbf{x}_t$, for brevity.
This is the general transformation formula for 
time derivative of vector field 
$\mathbf{a}$ between $L_I$ and $\hat{L}_R$.

A special case of eq.~(\ref{eq:c0021}) is given
when the vector field $\mathbf{a}$ is
Galilean invariant, i.e., $G=1$, such as the magnetic field $\mathbf{B}$ [see eq.~(\ref{eq:c0004})];
\begin{equation} \label{eq:c0023}
\frac{\partial \hat{\mathbf{B}}}{\partial t}(\hat{\mathbf{x}},t) 
= 
   R^{\Omega t}\, 
   [
       \frac{\partial}{\partial t}
        + (\mathbf{\Omega}\times\mathbf{x})\cdot \nabla
        - \mathbf{\Omega} \times
   ]
   \,        
    \mathbf{B}(\mathbf{x},t).
\end{equation}

The transformation of the fluid flow $\mathbf{u}(\mathbf{x},t)$
between $L_I$ and $\hat{L}_R$ is obtained by
substituting the Galilean transformation operator $G^\mathbf{V}$ for $\mathbf{u}$
defined in eq.~(\ref{eq:c0003}) into eq.~(\ref{eq:c0021});
\begin{align} \label{eq:c0024}
& \frac{\partial \hat{\mathbf{u}}}{\partial t}(\hat{\mathbf{x}},t)  \nonumber\\
&= 
   R^{\Omega t}\, 
   [
       \frac{\partial}{\partial t}
        + (\mathbf{\Omega}\times\mathbf{x})\cdot \nabla
        - \mathbf{\Omega} \times
   ]\,
       \mathbf{u}(\mathbf{x},t)  \nonumber\\
& +
   R^{\Omega t}
   [
         (\mathbf{\Omega}\times\mathbf{x})\cdot \nabla
        - \mathbf{\Omega} \times
   ]\, (\mathbf{x}\times\mathbf{\Omega}) \nonumber\\
& = 
   R^{\Omega t}\, 
   [
       \frac{\partial}{\partial t}
        + (\mathbf{\Omega}\times\mathbf{x})\cdot \nabla
        - \mathbf{\Omega} \times
   ]\,
       \mathbf{u}(\mathbf{x},t),
\end{align}
where we have used eq.~(\ref{eq:c2249}) 
with $\mathbf{a}=\mathbf{\Omega}\times\mathbf{x}$.
It is interesting that
the transformation rule of
the fluid flow $\mathbf{u}$ is exactly the same as that
of the magnetic field $\mathbf{B}$ 
although $\mathbf{u}$ is not a Galilean invariant vector.

%----------------------------
\section{Transformations of the Navier-Stokes equation}
%----------------------------

The Navier-Stokes equation for an incompressible fluid in
the inertial frame $L_I$ is written as
\begin{equation}  \label{eq:c2136}
  \frac{\partial \mathbf{u}}{\partial t}  
  + (\mathbf{u}\cdot \nabla) \mathbf{u} 
  =  \mathbf{f},
\end{equation}
where $\mathbf{f}\equiv -\nabla p + \nu \nabla^2\mathbf{u}$ with viscosity $\nu$.
The pressure gradient term $\nabla p$ is Galilean invariant vector,
and another term in the total force $\mathbf{f}$ is transformed as
\begin{eqnarray}   \label{eq:c2159}
  \nabla^2 \hat{\mathbf{u}}
   & = & 
   R^{\Omega t} G^{\mathbf{\Omega}\times\mathbf{x}}\, \nabla^2 \mathbf{u}  \nonumber\\
   & = & 
   R^{\Omega t}  \nabla^2 G^{\mathbf{\Omega}\times\mathbf{x}} \, \mathbf{u} 
   \ \ \ \hbox{[cf. eq.~(\ref{eq:c2307})]}\nonumber\\
   & = & 
   R^{\Omega t} 
   \nabla^2 (\mathbf{u}+\mathbf{x}\times\mathbf{\Omega})
    \ \ \ \hbox{[cf. eq.~(\ref{eq:c0014})]}    \nonumber\\
    & = & 
   R^{\Omega t} 
   \nabla^2 \mathbf{u}.
\end{eqnarray}
Therefore, 
the force term $\mathbf{f}$ is transformed 
as a Galilean invariant field;
\begin{equation}  \label{eq:c2150}
    \hat{\mathbf{f}}(\hat{\mathbf{x}},t) = R^{\Omega t}\, \mathbf{f}(\mathbf{x},t).
\end{equation}

Now, let us derive the transformed form of the Navier-Stokes equation~(\ref{eq:c2136}) in
the rotating frame $\hat{L}_R$.
Combining eqs.~(\ref{eq:c2058}) and~(\ref{eq:c0024}), we get
\begin{equation}  \label{eq:c2304}
  \frac{\partial \hat{\mathbf{u}}}{\partial t} 
  + (\hat{\mathbf{u}}\cdot\hat{\nabla})\hat{\mathbf{u}}
  = R^{\Omega t} 
  [
    \frac{\partial \mathbf{u}}{\partial t} + 
    (\mathbf{u}\cdot\nabla)\mathbf{u}
    + 2\mathbf{u}\times\mathbf{\Omega}
    +(\mathbf{x}\times\mathbf{\Omega})\times\mathbf{\Omega}
  ].
\end{equation}
The last two terms in the right-hand side are rewritten as follows
\begin{align} \label{eq:c2306}
& R^{\Omega t}
  \{
   2\mathbf{u}\times\mathbf{\Omega}
   + (\mathbf{x}\times\mathbf{\Omega})\times\mathbf{\Omega}
  \}  \nonumber\\
& =
  2R^{\Omega t} (\mathbf{u}\times\mathbf{\Omega})
  + R^{\Omega t} \{ (\mathbf{x}\times\mathbf{\Omega})
     \times\mathbf{\Omega} \} \nonumber\\
& =
  2 (R^{\Omega t}\mathbf{u}) \times\mathbf{\Omega}
  + \{ (R^{\Omega t} \mathbf{x})\times\mathbf{\Omega}\}
     \times\mathbf{\Omega} 
     \nonumber\\
& = 
 2\{ \hat{\mathbf{u}} - R^{\Omega t}
 (\mathbf{x}\times\mathbf{\Omega})\} 
  \times\mathbf{\Omega}  \ \ \ \ \hbox{[cf. eq.~(\ref{eq:c0014})]} \nonumber\\
& 
 \ \ \   + \{ (R^{\Omega t} \mathbf{x})\times\mathbf{\Omega}\}
     \times\mathbf{\Omega}   \nonumber\\
& = 
   2\hat{\mathbf{u}}\times\mathbf{\Omega}
   + (\mathbf{\Omega}\times\hat{\mathbf{x}})\times\mathbf{\Omega}
    \ \ \ \hbox{[cf. eq.~(\ref{eq:c0006})]},
\end{align}
which leads to
\begin{equation}
 \label{eq:c2311}
  \frac{\partial \hat{\mathbf{u}}}{\partial t} 
  + (\hat{\mathbf{u}}\cdot\hat{\nabla})\hat{\mathbf{u}}
  - 2\hat{\mathbf{u}}\times \mathbf{\Omega}
  - (\mathbf{\Omega}\times\hat{\mathbf{x}})\times\mathbf{\Omega}
  = R^{\Omega t} 
  [
    \frac{\partial \mathbf{u}}{\partial t} + 
    (\mathbf{u}\cdot\nabla)\mathbf{u}
  ].
\end{equation}
From eqs.~(\ref{eq:c2136}), (\ref{eq:c2150}), and (\ref{eq:c2311}),
we finally get
\begin{equation}  \label{eq:c2315}
  \frac{\partial \hat{\mathbf{u}}}{\partial t} 
  + (\hat{\mathbf{u}}\cdot\hat{\nabla})\hat{\mathbf{u}}
  = \hat{\mathbf{f}}
  + 2\hat{\mathbf{u}}\times \mathbf{\Omega}
  + (\mathbf{\Omega}\times\hat{\mathbf{x}})\times\mathbf{\Omega}.
\end{equation}
The second term in the right hand side is the Coriolis force,
and the third term is the centrifugal force.
Note that even if we 
have another Galilean invariant force
in $\mathbf{f}$ in the 
right hand side of eq.~(\ref{eq:c2136}), e.g., the buoyancy force,
the equation form of~(\ref{eq:c2315}) does not change.
Note also that 
the Lorentz force in the magnetohydrodynamics~\cite[e.g.,][]{davidson_2001}
is Galilean invariant.

%----------------------------
\section{Summary and Discussion}
%----------------------------
In this paper, we have shown a general algorithm to derive
the evolving equation in a constantly rotating frame of reference
based on the local Galilean transformation and the rotating
coordinate transformation of field quantities.
This derivation---Eulerian derivation---is 
applied in a straightforward way to a rotating fluid system to derive
the Coriolis force and the centrifugal force.
When the angular velocity of the rotating frame is time dependent,
i.e., $\dot{\mathbf{\Omega}}(t) \equiv d\mathbf{\Omega}(t)/dt \ne 0$,
the instantaneous local reference frame fixed at a position
$\mathbf{x}$ in the rotating system is not an inertial frame, but
rather an accelerating frame with the acceleration rate
$\mathbf{A}(\mathbf{x},t) = \dot{\mathbf{\Omega}}(t)\times\mathbf{x}$.
Therefore, another pseudo force, or the inertial force,
$-\mathbf{A}(\mathbf{x},t)$, per unit mass
appears in the equation of motion.
This term plays important roles in some rotating fluid dynamics
such as precession and nutations~\citep{greenspan_1990}.

The usefulness of the Eulerian derivation becomes evident when we apply
it to the derivation of the basic equation 
in the rotating system 
for other physical systems rather than the fluid.
Take the magnetic field $\mathbf{B}$ for the example.
From the Maxwell's equations, the induction equation of $\mathbf{B}$ is written as
\begin{equation} \label{eq:c2233}
  \frac{\partial \mathbf{B}}{\partial t}(\mathbf{x},t) = - \nabla \times \mathbf{E}(\mathbf{x},t),
\end{equation}
in the inertial frame $L_I$.
The $\nabla \times \mathbf{E}$ term in the 
right hand side of this equation is transformed into the following form in
the rotating frame $\hat{L}_R$:
\begin{align} \label{eq:c2234}
&  \hat{\nabla} \times \hat{\mathbf{E}}(\hat{\mathbf{x}},t)   \nonumber\\
&=  R^{\Omega t}\,
            G^{\mathbf{\Omega}\times\mathbf{x}}\,
             \nabla \times \mathbf{E}(\mathbf{x},t) 
             \ \ \ \hbox{[cf. eq.(\ref{eq:c0013})]}   \nonumber\\
&=  R^{\Omega t}\,
             \nabla \times 
                [ 
                    G^{\mathbf{\Omega}\times\mathbf{x}}\,\mathbf{E}(\mathbf{x},t)
                ]    \ \ \ \hbox{[cf. eq.(\ref{eq:c2307})]}  \nonumber\\             
& =  R^{\Omega t}\,
             \nabla \times 
                [
                                    \mathbf{E}(\mathbf{x},t) + (\mathbf{\Omega}\times\mathbf{x})\times\mathbf{B}
                ] \ \ \ \hbox{[cf. eq.(\ref{eq:c0016})]}   \nonumber\\
& =  R^{\Omega t}\,
            [ \nabla \times \mathbf{E}(\mathbf{x},t)
                                 -\{ 
                                      (\mathbf{\Omega}\times\mathbf{x})\cdot\nabla
                                  \} \mathbf{B}
                                 + \mathbf{\Omega}\times\mathbf{B}
                ].
\end{align}
Here we have used $\nabla\cdot\mathbf{B}=0$,
$\nabla\cdot(\mathbf{\Omega}\times\mathbf{x})=0$,
and eq.~(\ref{eq:c2249}) in the last step.
Comparing eq.~(\ref{eq:c0023}) with eq.~(\ref{eq:c2234}), we get
\begin{equation} \label{eq:c2319}
  \frac{\partial \hat{\mathbf{B}}}{\partial t}(\hat{\mathbf{x}},t)
   = - \hat{\nabla} \times \hat{\mathbf{E}}(\hat{\mathbf{x}},t).
\end{equation}
Therefore, the induction equation does not change 
its form in the rotating frame of reference.
(There is no ``apparent induction'' term.)
This example clearly illustrates the advantage of Eulerian 
approach of the transformation of the basic equation 
in the rotating frame.

%% ------------------------------------------------------------------------ %%
%
%  ACKNOWLEDGMENTS
%
%% ------------------------------------------------------------------------ %%

\begin{acknowledgments}
The authors would like to thank Shinya Kakuta and 
Mayumi Yoshioka for useful comments.
This research was partially supported by the Ministry of 
Education, Science, Sports and Culture, Grant-in-Aid for 
Scientific Research (C), 17540404, 2005.
\end{acknowledgments}

%% ------------------------------------------------------------------------ %%
%
%  REFERENCE LIST AND TEXT CITATIONS
%
%% ------------------------------------------------------------------------ %%
%\bibliographystyle{agu04}
%\bibliography{without_mymemo}

\end{document}